\documentclass[showpacs,amsmath,amssymb,aps,prl,twocolumn]{revtex4}

\usepackage{graphicx}
\usepackage{dcolumn}
\usepackage{bm}

\begin{document}

\title{{\sl Ab initio} investigation of competing instabilities in
  ferroelectric perovskite PbTiO$_{3}$}

\author{Jacek C. Wojde\l,$^{1}$ Patrick Hermet,$^{2,3}$ Mathias P.
  Ljungberg,$^{1}$ Philippe Ghosez,$^{2}$ and Jorge \'I\~niguez$^{1}$}

\affiliation{$^{1}$Institut de Ci\`encia de Materials de Barcelona
  (ICMAB-CSIC), Campus UAB, E-08193 Bellaterra, Spain\\$^{2}$Physique
  Th\'eorique des Mat\'eriaux, Universit\'e de Li\`ege (B5), B-4000
  Li\`ege, Belgium\\$^{3}$Laboratoire Charles Coulomb (UMR CNRS 5221),
  Universit\'e Montpellier II, FR-34095 Montpellier C\'edex 5, France}

\begin{abstract}
We have developed first-principles models, based on a general
parametrization of the full potential-energy surface, to investigate
the lattice-dynamical properties of perovskite oxides. We discuss the
application of our method to prototypic ferroelectric PbTiO$_{3}$,
showing that its properties are drastically affected by a competition
between structural instabilities. Indeed, we confirm previous
indications that the destructive interaction between the ferroelectric
and antiferrodistortive (involving rotations of the O$_6$ octahedra)
soft modes shifts PbTiO$_3$'s Curie temperature by hundreds of
degrees. Our theory provides unique insight into this gigantic effect
and its dynamical character.
\end{abstract}







\maketitle

Perovskite oxides are a large family of materials of great fundamental
and applied interest. In many cases, the structural and
lattice-dynamical features of the compounds are critical to determine
their properties. Most notably, this is the case of ferroelectric (FE)
perovskites -- whose spontaneous polarization is usually the result of
a structural phase transition~\cite{ferro} -- and related compounds
such as magnetoelectric multiferroics -- whose properties can be
greatly enhanced by engineering the lattice response to external
fields~\cite{wojdel10}. First-principles studies of lattice-dynamical
phenomena are usually restricted to the limit of low temperatures, as
the inclusion of thermal effects requires large simulation boxes
(thousands of atoms to get a realistic description of most properties
of interest) and a good sampling of configuration space (tens of
thousands of Monte Carlo sweeps to compute statistical
averages). Hence, it is not yet feasible to tackle such problems from
first-principles directly.

The situation greatly improved with the development of effective
models that (1) capture in a general and mathematically simple form
the energy surface associated with the most relevant (i.e.,
lowest-energy) structural distortions and (2) whose parameters are
computed from density-functional theory (DFT). This so-called {\em
  effective Hamiltonian} approach has been successfully applied to FE
crystals like BaTiO$_{3}$ \cite{zhong94} and PbTiO$_3$
(PTO)~\cite{waghmare97}, solid solutions like
PbZr$_{1-x}$Ti$_{x}$O$_{3}$ (PZT) \cite{bellaiche00,kornev06}, and
even multiferroics like BiFeO$_{3}$~\cite{kornev07}. However, by
neglecting all but the lowest-energy degrees of freedom, this scheme
may introduce quantitative (e.g., in the description of thermal
expansion~\cite{tinte03}) and qualitative (e.g., whenever secondary
modes are important~\cite{dieguez11}) errors. Further, the approach
does not seem well suited to tackle structurally complex cases -- from
domain walls to heterostructures displaying interface-driven
phenomena~\cite{bousquet08} -- in which it may be difficult to
identify a small set of distortions that capture the main
effects. Shell models \cite{sepliarsky05} and other schemes
\cite{shin05} providing a full atomistic description have also been
derived from first-principles for these materials; however, such
approaches rely on specific forms of the interatomic potentials, which
may restrict their accuracy and generality.

To overcome these limitations, we have developed an aproach that
provides a full atomistic description of the materials while retaining
the general energy parametrization of the effective-Hamiltonian
scheme. Here we describe our first application -- to the study of
PTO's FE phase transition --, which illustrates the predictive power
and unique insights that the new method offers.

{\sl Methods}.-- Let us briefly describe our all-atom effective
Hamiltonians, which will be presented in detail
elsewhere~\cite{wojdel-unp}. As in the usual approach~\cite{zhong94},
we write the energy of the material as a Taylor series around a (cubic
perovskite) reference structure. Our Taylor series, though, is a
function of {\em all} the atomic degrees of freedom, not only the
lowest-energy ones. It is convenient to split our variables into
atomic displacements (generically denoted by $u$) and global strains
of the simulation box ($\eta$), so that we can write the energy as
\begin{align}
E(u,\eta) = & \, E_0 + E(u^{2}) + E(\eta^{2}) + E(\eta u^{2})
\nonumber \\ & + E(u^{3},u^{4},...) + E(\eta^{3},\eta^{4},...) \\ & +
E(\eta^{2}u^{2},\eta^{3}u^{2},...,\eta u^{3},...) \, . \nonumber
\end{align}
The first line in Eq.~(1) contains the harmonic terms in $u$ (i.e.,
the force-constant matrix $\boldsymbol{K}$) and $\eta$ (frozen-ion
elastic tensor), which can be readily obtained from most
first-principles codes; we computed them for PTO using
density-functional perturbation theory as implemented in {\sc
  abinit}~\cite{fn:calculations,abinit,WCGGA,rappe90}. Figure~1 shows
the dispersion curves for the $\kappa_{\boldsymbol{q}s}$ eigenvalues
of $\boldsymbol{K}$, which include a variety of structural
instabilities of the cubic phase (i.e., distortions with
$\kappa_{\boldsymbol{q}s}<$~0) and are exactly captured by our
model~\cite{fn:longrange}. The first line of Eq.~(1) also includes the
lowest-order non-zero $\eta$-$u$ couplings~\cite{fn:symmetry}, which
we obtain from the force-constant matrices computed for slightly
strained cells. Such a {\em strain-phonon} coupling term is the only
$\eta$-$u$ interaction considered in the usual effective Hamiltonians
of ferroelectrics.

\begin{figure}[t!]
 \includegraphics[width=\columnwidth]{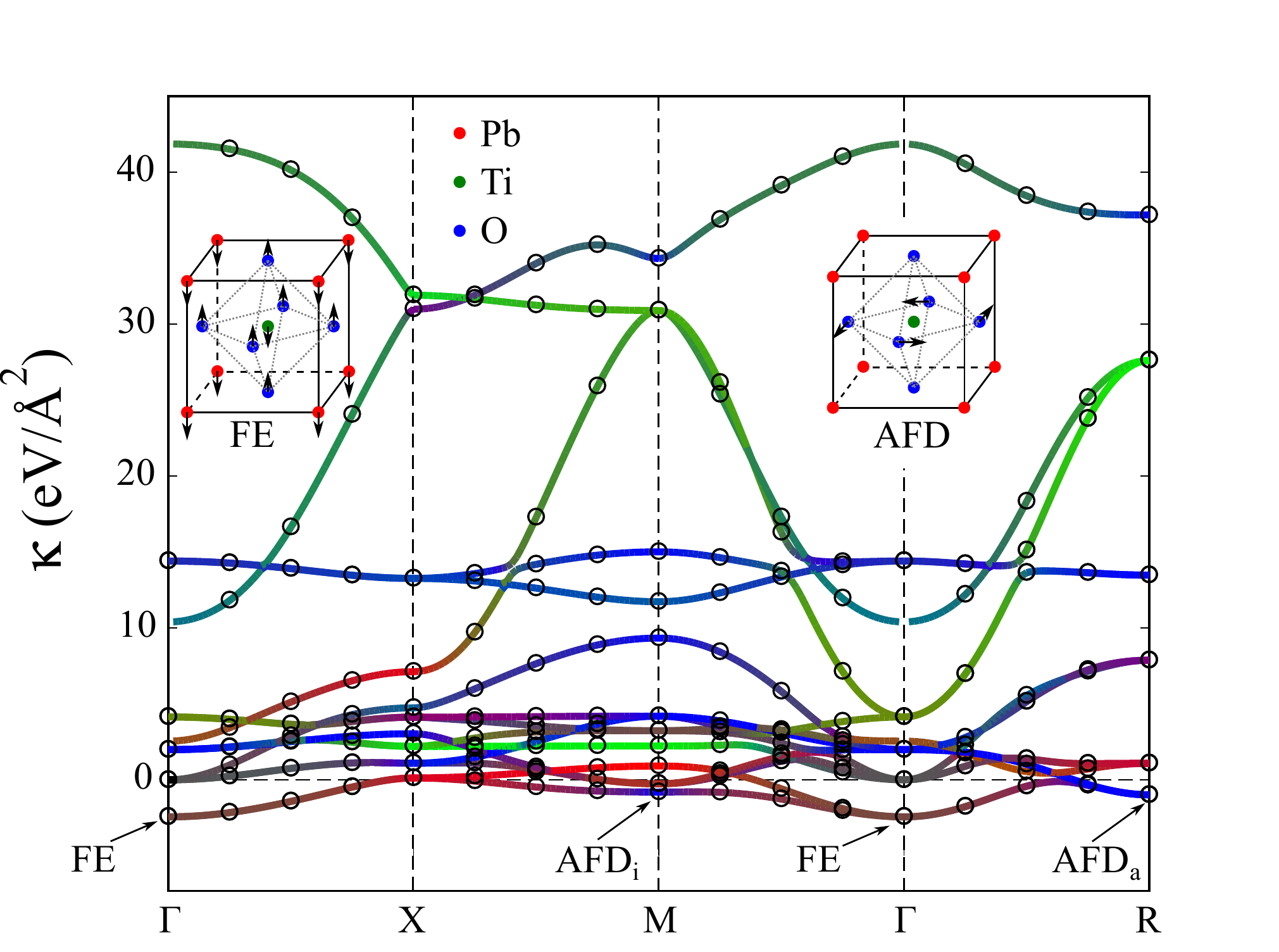}
 \caption{(Color online.) Dispersion curves for the
   $\kappa_{\boldsymbol{q}s}$ eigenvalues of the force-constant matrix
   $\boldsymbol{K}$ of PTO's cubic phase. Lines and circles are the
   results obtained {\sl ab initio} and with our effective model,
   respectively. The line color indicates the atomic character of the
   eigenmode; red, green, and blue correspond to Pb, Ti, and O,
   respectively. Unstable ($\kappa_{\boldsymbol{q}s} < 0$) FE and AFD
   modes are sketched. The AFD$_{\rm a}$ (AFD$_{\rm i}$) modes involve
   an antiphase (inphase) modulation of the O$_6$-rotations along the
   rotation axis.}
 \label{fig:phonons}
\end{figure}

\begin{figure*}[t!]
 \includegraphics[width=\textwidth]{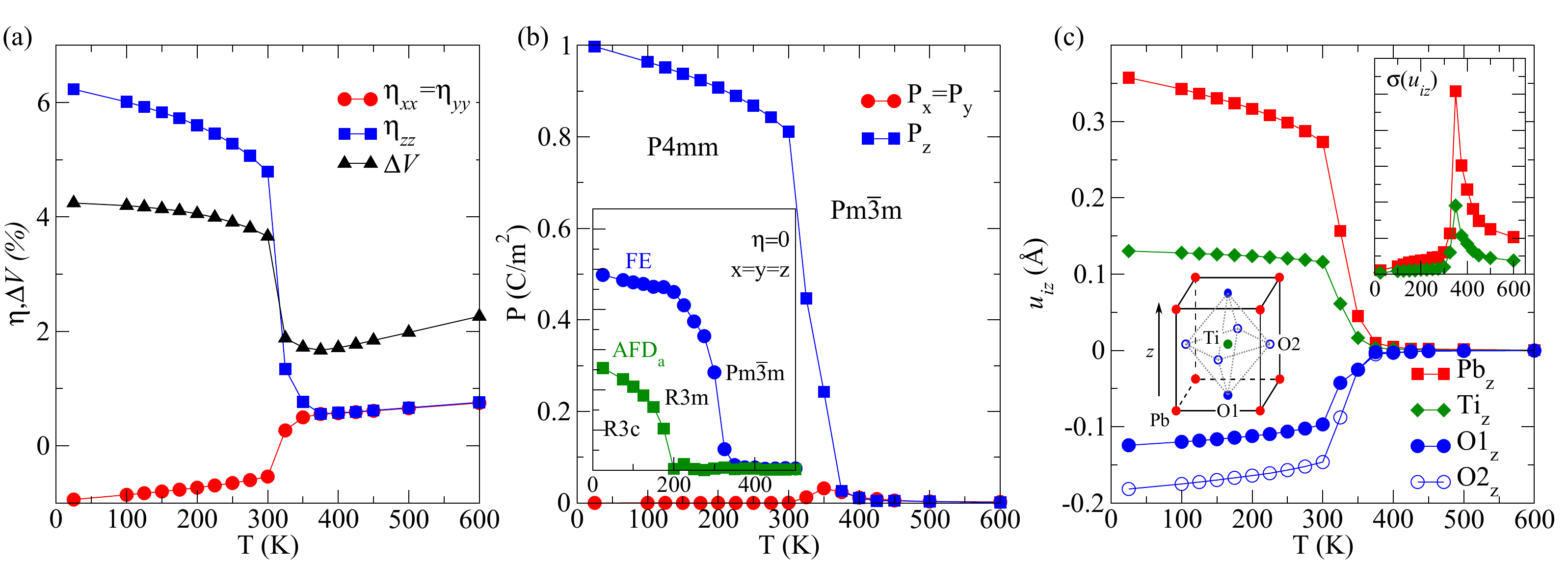}
 \caption{(Color online.) $T$-dependence of the quantities
   characterizing PTO's FE transition. Panel~(a): strains and volume
   change with respect to the reference cubic cell. Panel~(b):
   Spontaneous polarization $P_{\alpha} = v^{-1} \sum_{i\beta}
   Z_{i\beta,\alpha} u_{i\beta}$, where $v$ is the unit cell volume,
   $\alpha$ and $\beta$ are spatial directions, $i$ labels atoms
   within the unit cell, $Z_{i\beta,\alpha}$ are Born charges, and
   $u_{i\beta}$ are supercell-averaged atomic displacements. Inset:
   results of simulations at fixed $\eta$~=~0; here, the FE and
   AFD$_{\rm a}$ distortions are obtained by projecting the
   thermal-averaged configuration onto the corresponding unstable
   $\boldsymbol{K}$-eigenmodes of the cubic phase (see
   Fig.~\protect\ref{fig:phonons}); units are arbitrary. Panel~(c):
   Supercell-averaged displacements $u_{i\alpha}$ along the direction
   of the spontaneous polarization. A global displacement is chosen so
   that $\sum_{i} u_{i\alpha} = 0$. Inset: standard deviation of the
   Pb and Ti displacements, in arbitrary units.}
 \label{fig:t-scan}
\end{figure*}

The second and third lines of Eq.~(1) include anharmonic interactions
that in principle can extend to arbitrary order and need to be
truncated~\cite{fn:symmetry}. We computed the parameters in such terms
by fitting to the key features (energy, structure,
$\boldsymbol{K}$-eigenmodes at selected $\boldsymbol{q}$ points) of
some relevant low-symmetry low-energy phases, which include several
structures in which FE and antiferrodistortive (AFD) modes are
condensed. [As described in Fig.~1, AFD modes may involve inphase
  (AFD$_{\rm i}$) and antiphase (AFD$_{\rm a}$) rotations of the O$_6$
  octahedra; only the AFD$_{\rm a}$ modes are relevant in the present
  discussion.] We first considered low-symmetry phases obtained by
imposing the cubic cell ($\eta$~=~0) to compute the parameters in the
$E(u^{3},u^{4},...)$ term; we concluded it can be well approximated by
(1) truncating the series at fourth order and (2) including only
pairwise interactions between nearest-neighboring atoms (Pb-O and
Ti-O). We then considered the fully-relaxed low-symmetry structures,
and found that high-order $\eta$-$u$ couplings are needed in order to
describe their main features accurately and, simultaneously, reproduce
the first-order character of PTO's FE transition. (Our model includes
selected $\eta$-$u$ couplings up to $\eta^{4}u^{2}$ order.) The
$E(\eta^{3},\eta^{4},...)$ term, on the other hand, could be
neglected. The highest-order terms were chosen to be even powers of
$u$ and $\eta$, and the corresponding parameters restricted to be
positive, to guarantee the global stability of the model.

Typically, the effective model of PTO used in this work reproduces the
first-principles energies of the relevant low-symmetry phases within a
15\%, and their low-lying $\boldsymbol{K}$-eigenvalues within
0.5~eV/\AA$^2$. Note that our parameter-fitting procedure implicitly
captures the usual features of atomistic models of ferroelectrics. For
example, the double-well potential associated with the FE instability
stems from (1) the unstable polar modes included at the harmonic level
and (2) the anharmonic terms fitted to describe the stable
low-symmetry phases obtained {\sl ab initio}. The anharmonic
interactions between modes are captured by fitting to the
$\kappa_{\boldsymbol{q}s}$ spectrum of distorted structures, etc.

{\sl Basic results}.-- We solved our model by performing Monte Carlo
simulations, typically using 20000 sweeps for thermalization and 80000
for computing averages at each considered temperature ($T$). The
simulation box contained 8$\times$8$\times$8 perovskite cells (i.e.,
2560 atoms) and periodic boundary conditions were
employed. (10$\times$10$\times$10 supercells of 5000 atoms were used
for $T\approx T_{\rm C}$, where size effects become more relevant.)
Figure~\ref{fig:t-scan} shows our basic results. Our model predicts
that, at a temperature $T_{\rm C}\approx$~325~K, PTO undergoes a sharp
transition between the high-$T$ cubic ($Pm\bar{3}m$) and low-$T$
tetragonal ($P4mm$) phases. This transition carries with it the
deformation of the unit cell [Fig.~2(a)] and the development of an
spontaneous polarization $\boldsymbol{P}$ [Fig.~2(b)].

We confirmed that the polarization is the primary order parameter of
the transformation by running simulations with the cubic cell fixed
($\eta$~=~0); in that case we still obtain a FE transition at a
similar $T_{\rm C}$ [inset of Fig.~2(b)], but the polar phase has now
a rhombohedral symmetry ($R3m$) with $P_{x}=P_{y}=P_{z}$. Further, at
$T\approx$~200~K we observe a second transition in which rotations of
the O$_{6}$ octahedra around the polar axis are condensed (the
symmetry is $R3c$ and the cell doubles). These results show the
importance that the $\eta$-$u$ couplings have in determining PTO's
behavior, and suggest that mechanical constraints (e.g., epitaxial
strain) can greatly affect it.

Figure~2(c) shows the $T$-dependence of the atomic positions in PTO's
unit cell. We find that, while the magnitude of the FE distortion
depends strongly on $T$, the displacement pattern remains relatively
constant. Hence, in this sense our results validate the usual
assumption that the main features of the FE transition can be captured
by effective models including only one type of polar distortion. Note
also that at $T\approx T_{\rm C}$ the atomic displacements become {\em
  soft} [i.e., there is a large spread of their instantaneous values,
  as shown in the inset of Fig.~2(c)], which leads to an enhancement
of the dielectric and (below $T_{C}$) piezoelectric responses (not
shown here).

Because it includes all the atoms, our model should correct the usual
effective-Hamiltonian underestimation of the thermal
expansion~\cite{tinte03}. Indeed, above $T_{\rm C}$ we obtain an
expansion coefficient $\alpha$~=~$18.2\times
10^{-6}$~$^{\circ}$C$^{-1}$, which exceeds the experimental value of
$12.6\times 10^{-6}$~$^{\circ}$C$^{-1}$~\cite{haun87}.

{\sl What controls T$_{\rm C}$?}-- Our computed $T_{\rm C}$ is about
325~K, while the experimental result is around
760~K~\cite{haun87}. This severe underestimation is surprising, as
previous studies of PTO based on less-complete theories had shown a
much better agreement with experiment. In particular, Waghmare and
Rabe (WR) \cite{waghmare97} constructed a model that neglects all
degrees of freedom but the soft polar modes (treated as lattice
Wannier functions) and cell strains, and obtained a value of
660~K~\cite{fn:batio3}.

We checked our model reproduces the energetics of the FE instabilities
given by the WR Hamiltonian quite closely, despite the differences
(e.g., DFT functionals, pseudopotentials) in the {\sl ab initio}
calculations employed to compute the parameters. Further, we ran
simulations with modified versions of our model to test subtle
features of the WR energy parametrization (e.g., the inclusion of
high-order terms for the FE distortions), and concluded that they have
a negligible effect on $T_{\rm C}$.

We thus turned our attention to the qualitatively distinct features of
our model. Most notably, we describe not only the FE instabilities,
but also the unstable AFD distortions shown in Fig.~1. It is known
that, in most perovskite oxides, the interaction between FE and AFD
modes is a competitive one, so that they tend to suppress each
other. The best studied case may be that of SrTiO$_3$ (STO), for which
Zhong and Vanderbilt predicted, by means of an effective Hamiltonian
approach, that the temperature of STO's AFD-related structural
transition would be about 25\% higher if the soft FE distortions of
the material were suppressed~\cite{zhong95}.

\begin{figure}[t!]
 \includegraphics[width=\columnwidth]{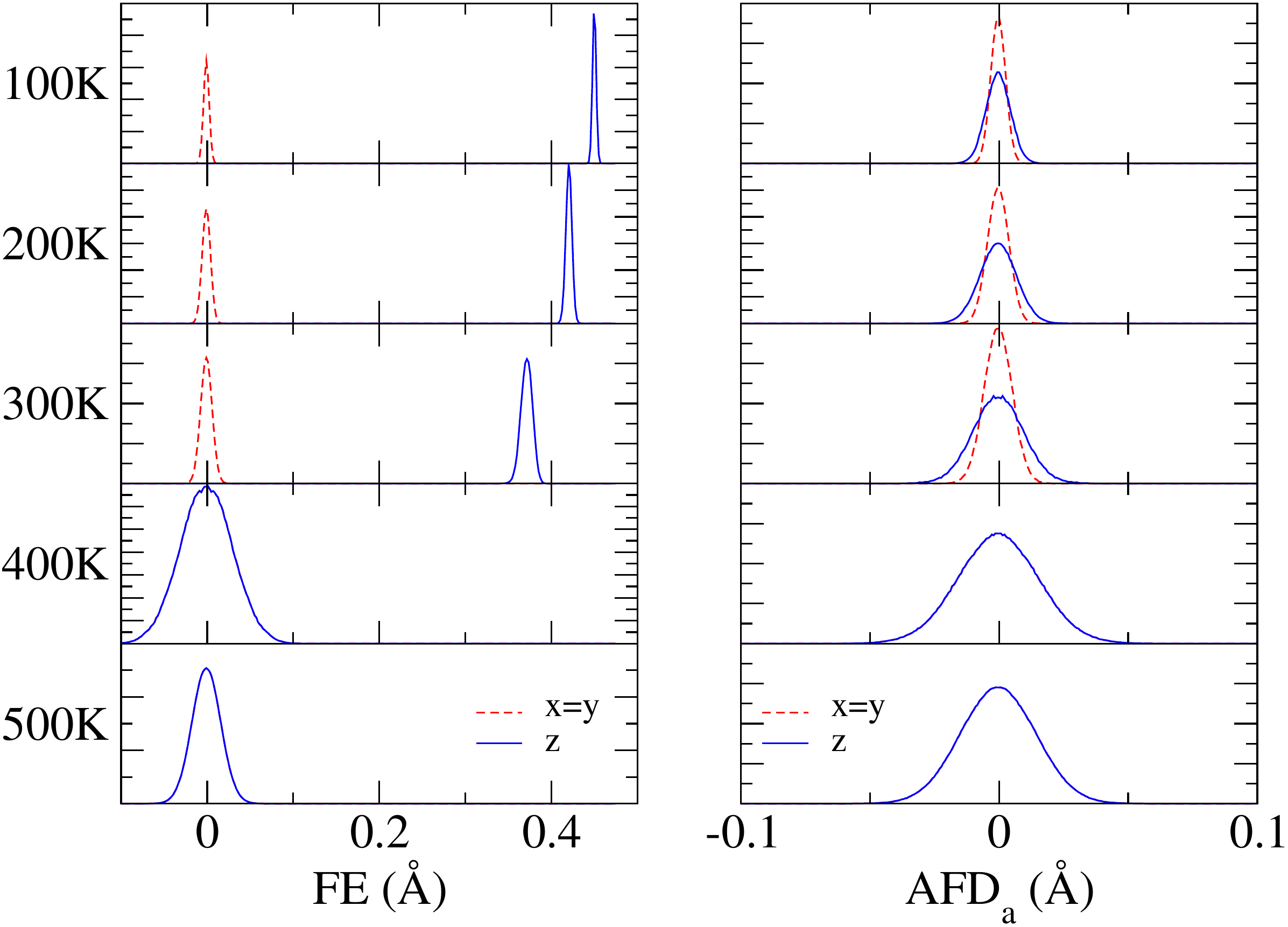}
 \caption{(Color online.) Histograms of the FE and AFD$_{\rm a}$
   distortions, which are quantified by projecting the instantaneous
   atomic structures on the corresponding unstable eigenmodes of the
   cubic reference phase.}
 \label{fig:histograms}
\end{figure}

The histograms in Fig.~\ref{fig:histograms} show that similar effects
occur in our PTO simulations.  Above $T_{\rm C}$, the AFD modes have
relatively large amplitudes, comparable in magnitude with those of the
FE distortions. In contrast, they get significantly suppressed below
$T_{\rm C}$, in a way that clearly reflects the breaking of the cubic
symmetry: the $z$-oriented spontaneous polarization restrains more
strongly the transversal O$_6$-rotational modes AFD$_{{\rm a}x}$ and
AFD$_{{\rm a}y}$. Note that the simultaneous occurrence of
FE$_{\alpha}$ and AFD$_{{\rm a}\beta}$ distortions, with
$\alpha\neq\beta$, results in a significant deformation of the O$_6$
octahedra, which may explain why these transversal FE-AFD interactions
are particularly destructive.

\begin{figure}[t!]
 \includegraphics[width=\columnwidth]{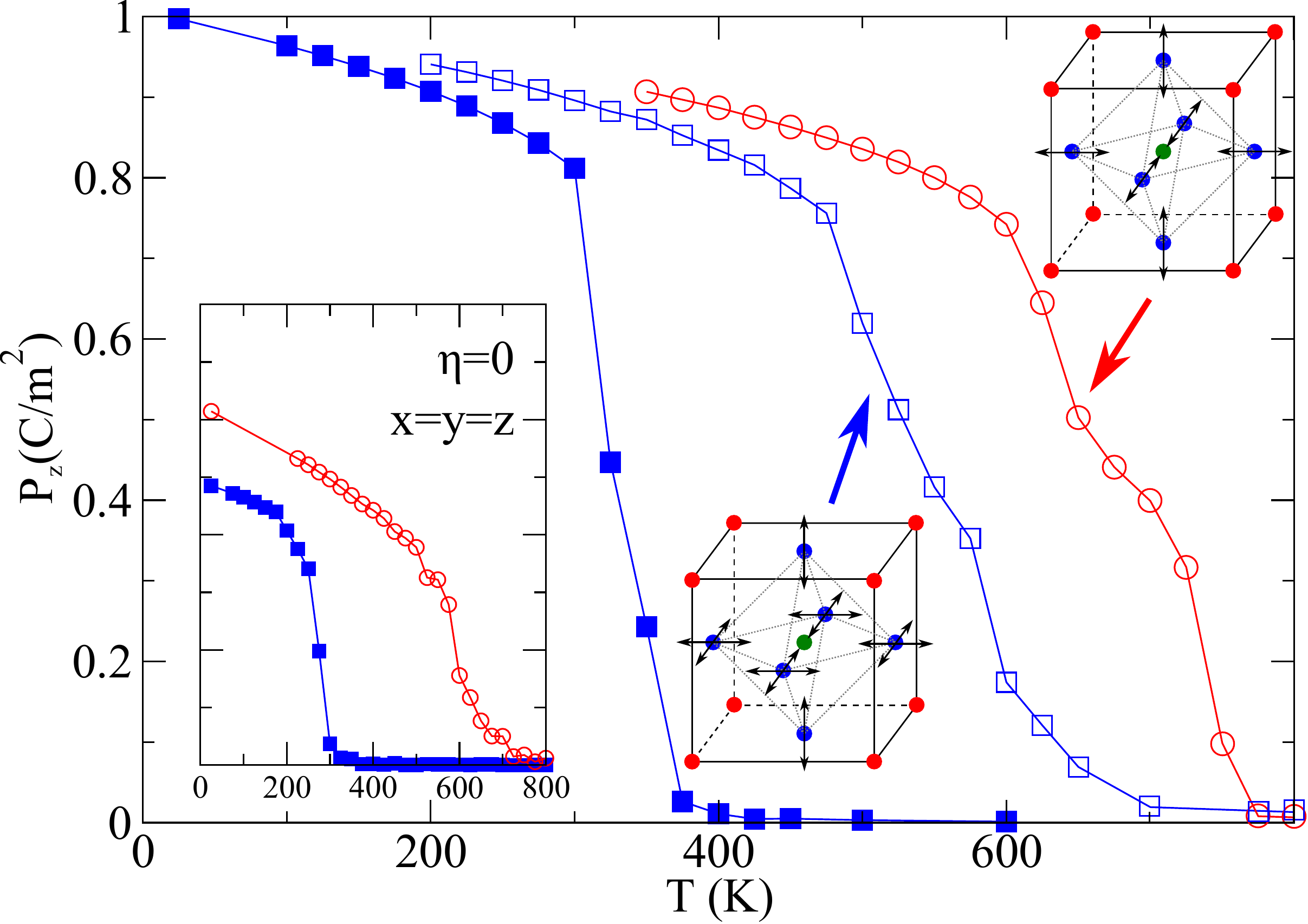}
 \caption{(Color online.) $T$-dependence of the polarization (only one
   component is shown) as obtained in the regular (solid squares),
   only-AFD$_{z}$ (open squares), and no-AFD (open circles) cases. The
   arrows in the sketches indicate the oxygen displacements that are
   allowed in the only-AFD$_{z}$ and no-AFD cases; the motion of the
   Pb and Ti atoms is unconstrained. Inset: Results from simulations
   with fixed $\eta$~=~0; we show the spontaneous polarization
   distortion for the regular (filled squares) and no-AFD (open
   circles) cases.}
 \label{fig:t-scan-fixedrotations}
\end{figure}

To investigate the effect of the AFD distortions on the FE transition
temperature, we ran simulations in which the O$_6$ rotations were
either totally ({\em no-AFD} case) or partly ({\em only-AFD$_{z}$}
case) suppressed. We imposed these constraints by restricting the
motion of the oxygen atoms as shown in the sketches in
Fig.~\ref{fig:t-scan-fixedrotations}. Let us stress that these
constraints do not affect the energetics associated with the
development of the spontaneous polarization, the FE ground state being
exactly retained. (Even in the only-AFD$_{z}$ case -- in which O$_6$
rotations are allowed only around the $z$ axis, which breaks the cubic
symmetry of the model --, we still keep the six equivalent FE minima
with $\boldsymbol{P}$ along the $\pm x$, $\pm y$, and $\pm z$
directions.)  Figure~\ref{fig:t-scan-fixedrotations} shows the
results: In the no-AFD case we obtain $T_{\rm C}\approx$~650~K, which
is very close to the WR result. We also ran this case at fixed
$\eta$~=~0 (inset of Fig.~\ref{fig:t-scan-fixedrotations}) and
obtained a similarly high $T_{\rm C}$. In the only-AFD$_{z}$ case the
calculated $T_{\rm C}$ is about 500~K, and the spontaneous
polarization is found to point specifically along $z$. (We checked
this by running several independent simulations starting from the
$u$~=~$\eta$~=~0 configuration.) Note that these effects -- i.e., the
shift in $T_{\rm C}$ and the preferential $\boldsymbol{P}$-direction
in the only-AFD$_{z}$ case -- have a strictly {\em dynamical}
origin. For example, in our only-AFD$_{z}$ simulations, the AFD$_{{\rm
    a}z}$ component fluctuates around its zero average value and
hampers the development of the polarization, especially along the
perpendicular directions $x$ and $y$. Thus, $P_{z}$ is {\em
  dynamically} favored and condenses at $T_{\rm C}$.

A couple of additional points are worth making. (1) We find that
strain does not play a big role in how the FE-AFD competition affects
$T_{\rm C}$, as our regular and fixed-cell simulations lead to very
similar results in this regard. Note that, in principle, one could
have expected otherwise: Because the FE and AFD instabilities tend to
react differently to an external
pressure~\cite{fn:pressure,janolin08,ganesh09}, a FE distortion should
imply cell deformations that, in turn, should suppress the AFD modes,
and {\sl viceversa}. However, our simulations suggest that such an
effect is not very important, probably because at the temperatures at
which the FE-AFD competition is relevant -- i.e., in the range from
350~K to 650~K where the material would be FE in absence of AFD modes
--, the correlations between atomic displacements and strains are
relatively small.

(2) Our claim that a destructive FE-AFD interaction exists in the
fixed-cell case may seem surprising, as a combined FE+AFD ground state
occurs for $\eta$~=~0 [see inset of Fig.~2(b)]. However, note that
competing instabilities may not fully suppress each other, and that is
the case here. To make the argument more quantitative, let us note
that the fixed-cell FE phase lies $-$59~meV/f.u. below the cubic
reference structure, while we get the AFD phase at $-$56~meV/f.u. The
combined FE+AFD structure has an energy of $-$62~meV/f.u., far above
the value of $-$115~meV/f.u. (where 115~=~59$+$56) that one would
expect if the FE and AFD modes did not interact at
all~\cite{fn:energy}. It is thus clear that these two instabilities
compete in the fixed-cell case, even if they coexist at low
temperatures. [In essence, in the regular case the polar distortion is
  larger (1.01~C/m$^2$ at 0~K, as compared with 0.57~C/m$^2$ when
  $\eta$~=~0) and the FE-AFD interaction fully suppresses the AFD
  instability.]

Let us note that a very large competition-driven shift in $T_{\rm C}$
had already been predicted (but not much emphasized) for PZT by Kornev
{\sl et al}.~\cite{kornev06}. The effective model used by these
authors included some particular FE-AFD couplings that allowed them to
rectify the very large $T_{\rm C}$ obtained previously from a simpler
theory~\cite{bellaiche00}. In contrast, our Hamiltonian captures the
FE-AFD couplings in an implicit and non-specific way; in fact, when
constructing our model we were not aware of the importance of this
interaction, nor did we need to assume any form for it. Hence, our
work ratifies that gigantic competition-driven effects can occur in
PTO-related compounds. Interestingly, the analogous phenomena in
SrTiO$_3$, which is generally considered the prototypic example for
competing structural instabilities, are probably much smaller; in
particular, Zhong and Vanderbilt obtained shifts in the transition
temperatures of about 35~K~\cite{zhong95}.

The origin of the large difference between our computed $T_{\rm C}$
(325~K) and the experimental one (760~K) remains an open question. We
ran additional tests to check the effect of some of the approximations
made in our model~\cite{wojdel-unp}, and did not observe any
significant improvements. Hence, we tend to attribute the disagreement
to the first-principles methods used to compute our Hamiltonian
parameters. Indeed, to obtain the right $T_{\rm C}$, the {\sl ab
  initio} simulations should describe accurately (1) the strength of
the FE instabilities, (2) the strength of the AFD instabilities, and
(3) the magnitude of the FE-AFD coupling. Our results suggest that
quantity (1) may be underestimated, and (2) and (3) overestimated, by
our current calculations, despite our using the best DFT methods
available for this task (i.e., recent functionals that have been shown
to perform very well for these
materials~\cite{bilc08,wahl08}). Clearly, $T_{\rm C}$'s large
sensitivity to the FE-AFD competition sets a very stringent
requirement for the accuracy of the {\sl ab initio}
calculations. Thus, our work suggests that, in spite of recent
progress, we still lack DFT methods that describe accurately the
thermodynamic properties of FE materials like PTO.

{\sl Summary}.-- We have introduced a novel approach for
first-principles investigations of the lattice-dynamical properties of
perovskite oxides. More precisely, we have developed effective models
that are based on a general parametrization of the energy of the
material (i.e., we make no {\sl a priori} assumptions on the form of
the interatomic couplings) and include all atomic degrees of
freedom. The application of the new scheme to PbTiO$_{3}$ has allowed
us to investigate the competition of instabilities at play in this
compound, which turns out to have dramatic effects in its
properties. Indeed, we found that such a competition results in a
reduction of PbTiO$_3$'s Curie temperature by as much as 300~K. Our
simulations provide unique insights into this gigantic effect and its
dynamical character.

Work supported by EC-FP7 project OxIDes (Grant No. CP-FP
228989-2). Also partly funded by MINECO-Spain (Grants
No. MAT2010-18113, No. MAT2010-10093-E, and No. CSD2007-00041) and
CSIC's JAE-doc program (JCW). Ph.G. thanks the Francqui Foundation for
Research Professorship. We acknowledge discussions with L.~Bellaiche,
A.~Garc\'{\i}a, and P.~Garc\'{\i}a-Fern\'andez.


\begin{thebibliography}{99}
\bibitem{fn:symmetry} Some terms are missing in Eq.~(1) -- e.g., those
  going as $\eta u$ or $\eta^2 u$ -- because they are zero by symmetry
  for the cubic perovskite structure.
\bibitem{fn:batio3} The $T_{\rm C}$ value that WR predicted for PTO is
  uncharacteristically close to experiment. For comparison, in the
  case of BaTiO$_{3}$ the effective Hamiltonian of Zhong {\sl et
    al}. \protect\cite{zhong94} predicts a $T_{\rm C}$ (about 300~K)
  that is 25\% smaller than the experimental result (about 400~K).
\bibitem{fn:pressure} Moderate compression tends to suppress the usual
  FE distortion and favor O$_6$ rotations. Such effects can be
  understood from steric arguments, and supporting evidence for
  specific materials can be found in the literature (e.g., for
  PbTiO$_3$ \protect\cite{janolin08,ganesh09} and
  SrTiO$_3$~\protect\cite{zhong95}).
\bibitem{fn:energy} For reference, the energy of the fully-relaxed
  tetragonal ground state is $-$123~meV/f.u. in our model.
\end{thebibliography}
\end{document}